\documentclass[epj]{webofc}
\usepackage[varg]{txfonts}   
\usepackage{graphicx}
\usepackage{epsfig}
\usepackage{subfigure}
\usepackage{amsmath}

\woctitle{New Frontiers in Physics 2014.}
\begin{document}
\title{Perspectives of Open Charm Physics at $\bar PANDA$}

\author{Elisabetta Prencipe\inst{1}\fnsep\thanks{\email{e.prencipe@fz-juelich.de}}
        on behalf of the $\bar PANDA$ Collaboration. 
}

\institute{Forschungszentrum J\"ulich, Leo Brandt strasse, 52428 J\"ulich, Germany}

\abstract{
 The  $\bar PANDA$ experiment at FAIR (Facility for Antiproton and Ion Research) in Darmstadt (Germany) is designed for $\bar p p$ annihilation studies and it will investigate fundamental questions of hadron and nuclear physics in interactions of antiprotons with nucleons and nuclei. Gluonic excitations and the physics of hadrons with strange and charm quarks will be accessible with unprecedented accuracy, thereby allowing high precision tests of the strong interactions. In particular, the $D_{s0}^*(2317)^+$ and $D_{s1}(2460)^+$ are still of high interest 11 years after their discovery, because they can not be simply understood in term of potential models. The available statistics and resolution of the past experiments did not allow to clarify their nature. Recently LHCb at CERN has made progresses in this respect, but still not at the level of precision required in order to clarify the puzzle of the $cs$-spectrum. $\bar PANDA$ will be able to achieve a factor 20 higher mass resolution than attained at the B-factories, which is expected to be decisive on these and second-order open questions. The technique to evaluate the width from the excitation function of the cross section of the $D_s$ mesons will be presented, and ongoing simulations performed with $PandaRoot$ will be shown.
}
\maketitle
\section{Introduction}
\label{intropanda}
The modern theory of strong interactions is the Quantum Chromo Dynamics (QCD), which is now well tested at high energy scale, but at low energy open questions remain, since only in the high energy regime  the coupling constant $\alpha_s$ is small and standard perturbative methods can be  applied. The sector of Charm and Charmonium physics is richer than expected respect to the quark model, as new resonant states with quite unusual properties have been observed. Prominent examples are the $X(3872)$ and the charged $Z_c^+(3900)$ in the Charmonium sector, and the $D_s$ mesons below the $DK$ threshold in the Charm sector. 
Several theoretical interpretations have been proposed for the new resonant states like hadro-charmonia, hybrids, tetraquarks and hadronic molecules. A recent review can be found at Ref.~\cite{review-panda1, review-panda2, review-panda3}. High quality calculations as well as measurements are compulsory for each state to allow one to decide amongst the various scenarios. 

\begin{figure*}[ht] 
\begin{center}
\mbox{
\subfigure[]{\scalebox{0.4}{\includegraphics{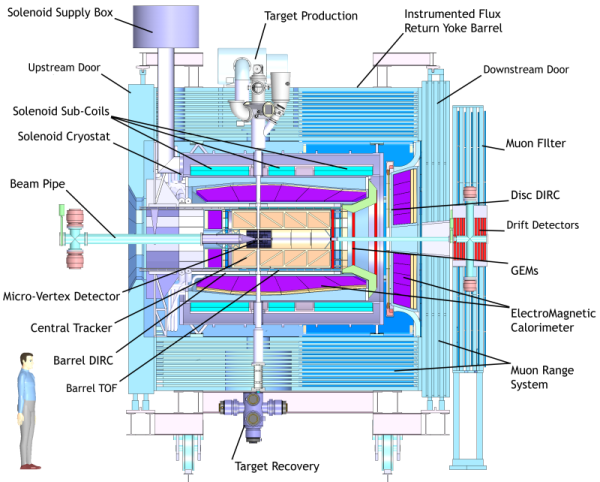}}}\hspace{-4.8 mm} \quad
\subfigure[]{\scalebox{0.2}{\includegraphics{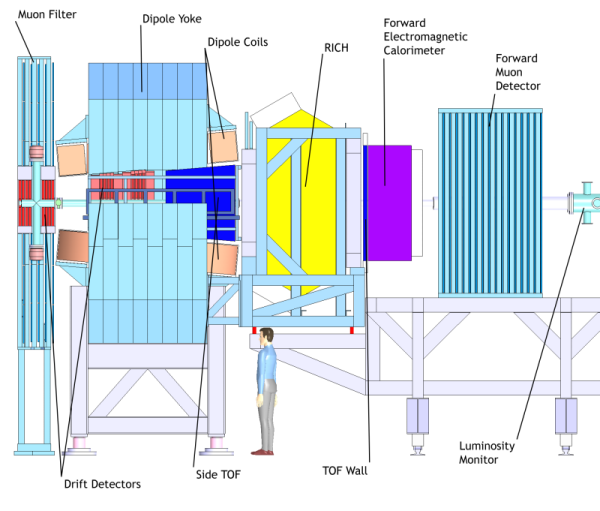}}}\quad
\subfigure[]{\scalebox{0.4}{\includegraphics{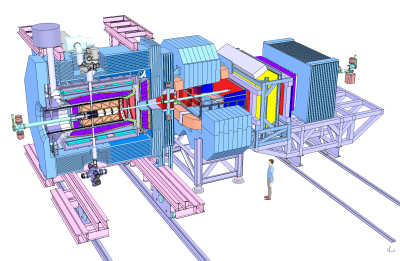}}} \hspace{130 mm}
}
\caption{\label{Fig10-panda} Detailed view of the (a) central spectrometer and (b) forward spectrometer of $\bar PANDA$; (c) general view of the $\bar PANDA$ detector.}
\end{center}
\end{figure*}

$\bar PANDA$ (antiProton ANnihilation at DArmstadt) at FAIR is designed as a fixed target machine, composed by two main parts: the central spectrometer, inserted in a  homogeneous magnetic field ($B$ = 2T); and the forward spectrometer in a dipole field ($B$ = 2T$\cdot$m),  as shown in Fig.~\ref{Fig10-panda}. High performant tracking and particle identification devices are designed for this experiment. $\bar PANDA$ will span a wide momentum range, from 1.5 up to 15 GeV/c. Focalized through stochastic and electron cooling, the antiproton beam will have excellent momentum resolution. This allows to challenge an ambitious physics program in nuclear and particle physics, detailed explained in Ref. \cite{pandabook}. Charm and Charmonium specroscopy are highlights of this program.

 This report focuses on experimental and theoretical aspects in the context of the exotic states $D_{s0}^*(2317)^+$ and $D_{s1}(2460)^+$, and what the original contribution of the future $\bar PANDA$ experiment will be in this respect. 
The status of the current Monte Carlo (MC) simulations is shown in this report.

\section{Theoretical approach}
\label{thpanda}
The study of charm physics is important from the point of view of strong and weak interactions.  Gluonic excitations and hadrons composed by strange and charm quarks can be abundantly produced in $\bar p p$ interactions and their features will be accessible with unprecedented accuracy, thereby allowing high precision tests of the strong interaction theory in the intermediate energy regime. On the other hand, searching for CP violation in the D sector recently has gained more attention, as new field of investigation.
\begin{figure*}[ht] 
\begin{center}
\mbox{
\subfigure[]{\scalebox{0.28}{\includegraphics{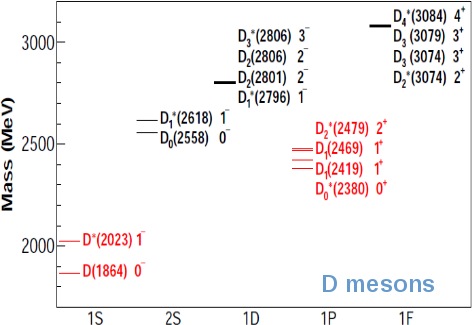}}}\quad
\subfigure[]{\scalebox{0.28}{\includegraphics{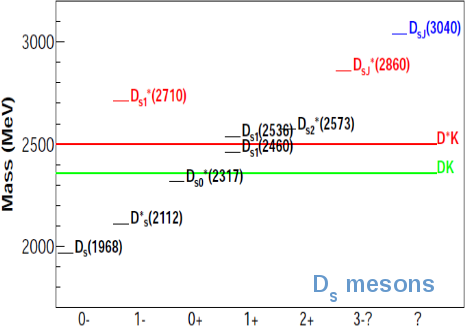}}}
}
\caption{\label{Fig0-panda}  Mass spectrum versus the particle spin J$\rm ^{PC}$ for (a) D mesons and  (b) D$_s$ mesons. The horizontal long lines indicate the DK and D$^*$K threshold; the short horizontal lines indicate the predicted states as in Ref.~\cite{dipierro1, dipierro2}. The mass spectra here reported are built using recent experimental results, presented at the conference CHARM2013 and available in Ref.~\cite{charm2013}.}
\end{center}
\end{figure*}

Understanding the $cs$-spectrum is not easy (see Fig.~\ref{Fig0-panda}): 11 years after the discovery of the charged state called $D_{s0}^*(2317)$\cite{Dsbabar}, its mass is known with high precision, but for its width only an upper limit exists. The observation of the  $D_{s0}^*(2317)$ represents a break-point, because the existence of combined $c$,$s$ quark systems is theoretically predicted\cite{dipierro1, dipierro2}; but some experimental observations questioned the potential models, which fairly agree with the observed D meson spectrum, e.g. mesons composed of the light quark $u$ or $d$ and the heavy quark $c$. Potential models agree also with the observation of several $D_s$ states, up to the discovery of the $D_{s0}^*$(2317).  But they cannot explain why the $D_{s0}^*(2317)$ mass was observed more than 100 MeV/c$^2$ below the predictions. The same is valid for the observation of the so called $D_{s1}(2460)$\cite{Ds2460}: its mass was found below the theoretical expectation as well. $D_{s0}^*(2317)$ and $D_{s1}(2460)$ are both very narrow, and  it is difficult to predict the cross section of $\bar p p \rightarrow D_s^{+(*)} D_s^{-(*)}$, as we cannot perform perturbative calculations, since they would underestimate the real cross section. The quantum numbers of $D_{s0}^*(2317)$ and $D_{s1}(2460)$ are, in any case, not fixed yet, although we can exclude $J{^P}$=0$^+$ for the $D_{s1}(2460)$ based on experimental observations\cite{widthgg}.

\begin{figure*}[ht] 
\begin{center}
\mbox{
\subfigure[]{\scalebox{0.2}{\includegraphics{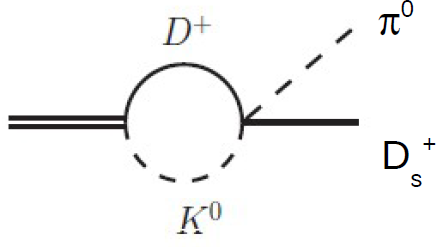}}}\quad
\subfigure[]{\scalebox{0.2}{\includegraphics{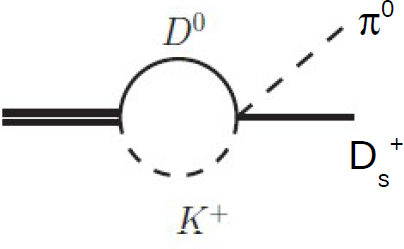}}}\quad 
\subfigure[]{\scalebox{0.27}{\includegraphics{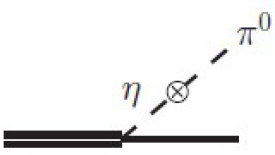}}} \vspace{-50mm}
}
\caption{\label{Fig2-panda} The two mechanisms that contribute to the hadronic width of the charged state $D_{s0}^*(2317)$\cite{christoph}: they represent the non-vanishing difference for the loop with (a) $D^+K^0$ and (b) $D^0K^+$, respectively. (c) depicts the decay via $\pi^0 - \eta$ mixing. If the $D_{s0}^*(2317)$ is a molecular states, the contributes (a) and (b) are not cancelling each other, and they add an important contribution to the total calculation of the hadronic width of the $D_{s0}^*(2317)$.}
\end{center}
\end{figure*}

The $D_{s0}^*(2317)$ mass is located by about 100 MeV/c$^2$ below what is predicted by the quark model and only about 40 MeV below the $DK$ threshold. While the state is an isoscalar, it can only decay strongly into the isovector final state $\pi^0 D_s$ which results in a width well below 1 MeV. Within the former explanations, the width of the state is predicted to be of the order of 10 keV\cite{purecs}; but the molecular scenario predicts consistently a width of the order of or even larger than 100 keV, if the enhancement stems from meson loops (see Fig.~\ref{Fig2-panda}), which are prominent for molecular states only\cite{ref5,ref6,ref7,ref8,ref9}. Measuring that small widths is clearly a challenge, but certainly worth the effort for such a measurement. It could for the first time unambiguously identify an exotic structure in the open charm sector.
\section{Experimental approach}
\label{sec-2}

$\bar PANDA$  can take the challenge to measure  the width of resonant states. The project $\bar PANDA$ aims to reach a mass resolution of 100 keV, which is 20 times better than attained at the B factories, and more than 2 times better than that of the Fermilab experiment E760. The reason is that it depends only on the beam resolution, that is expected to be $\Delta p$/$p$= 4$\cdot$10$^{-5}$.
\begin{figure*}[ht] 
\begin{center}
\mbox{
\subfigure[]{\scalebox{0.17}{\includegraphics{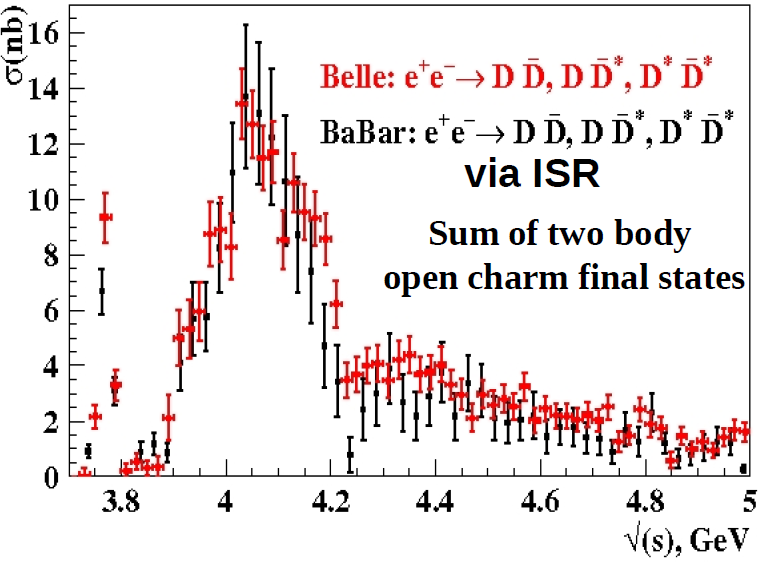}}}
\subfigure[]{\scalebox{0.15}{\includegraphics{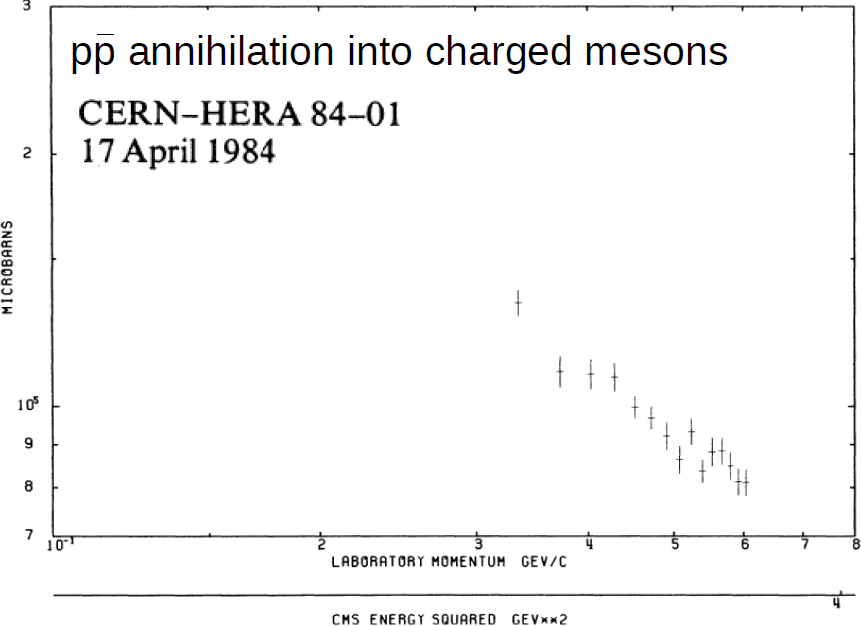}}}\quad
\subfigure[]{\scalebox{0.16}{\includegraphics{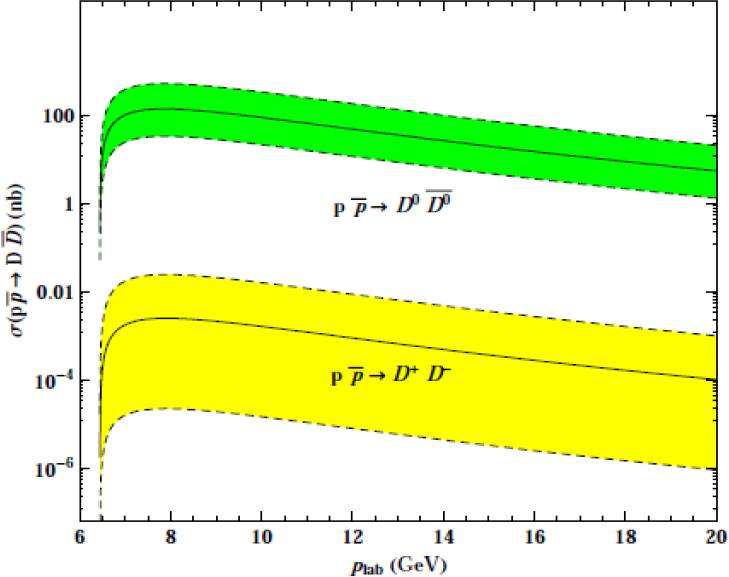}}}
}
\caption{\label{Fig1-panda} Cross sections evaluated at (a) B-factories\cite{crossBfactory1, crossBfactory2} and (b) at CERN\cite{crossCERN}. Cross section predictions on $\bar p p \rightarrow D \bar D$ are detailed explained in~\cite{mannel-panda}, but only for the D meson ground states (c): the presence of the quark $s$ in $D_s$ mesons adds much higher level of complications in these calculations, especially for the excited $D_s$ states.}
\end{center}
\end{figure*}

In the past 2 years the experiment LHCb improved the knowledge of  the $D_s$ spectrum, and confirmed the previous measurements, with the highest world precision; but it cannot provide the measurement of the width of these very narrow states, so the nature of the $D_s$ excited states is unclear. With the mass scan techniques in steps of 100 keV, $\bar PANDA$  is in a unique position to perform the study of the excitation function of the cross section (see Eq.~\ref{eq1-panda}), and discriminate among several theoretical models.  The cross section at given energy $\lambda$ is given by:
\begin{equation}
\label{eq1-panda}
\sigma(\lambda) = \sqrt{m_R \Gamma}  \cdot \lvert \mathcal{ M} \rvert^{\rm 2}  \cdot \frac{1}{\pi} \int_{- \infty}^{\lambda} \frac{\lambda -x}{x^{2} - 1} dx , 
\end{equation}
\begin{equation}
\label{eq2-panda}
\sigma(0) = \sqrt{m_R \Gamma / {\rm 2}} \cdot \lvert \mathcal{ M} \rvert^{\rm 2}
\end{equation}
where $\mathcal{M}$ is the matrix element, $m_R$ is the resonance mass and $\Gamma$ its width, $\lambda$= ($\sqrt s - m_R - m_{Ds}$)/$\Gamma$, $\sqrt s$ = energy in the center of mass for the production e.g. of $D_{s0}^*(2317)$ in the process $\bar p p \rightarrow D_s^- D_{s0}^{*}$(2317)$^+$,  that is equal to 4.286 GeV/c$^2$ at threshold of this process, with beam momentum $p_b$ = 8.8 GeV/c.
\begin{figure*}[ht] 
\begin{center}
\mbox{
\subfigure[]{\scalebox{0.28}{\includegraphics{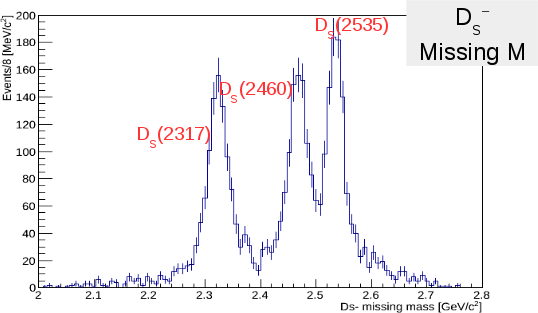}}}\quad
\subfigure[]{\scalebox{0.28}{\includegraphics{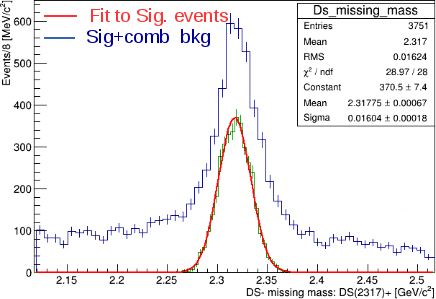}}}\quad
\subfigure[]{\scalebox{0.355}{\includegraphics{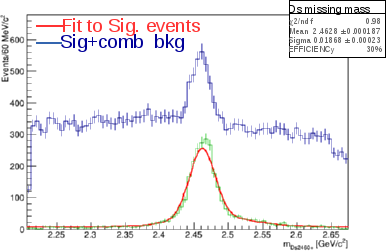}}}
}
\caption{\label{Fig4-panda} (a) $\bar PANDA$ MC simulations of the physics process $\bar p p \rightarrow D_s^- D_{sJ}^+$, where $D_{sJ}^+$ = $D_{s0}^{*}$(2317)$^+$, $D_{s1}$(2460)$^+$ and $D_{s1}'$(2536)$^+$, with $D_{s0}^{*}$(2317)$^+ \rightarrow D_s^+ \pi^0$, $D_{s1}$(2460)$^+$ $\rightarrow D_s^{*+} \pi^0$, $D_{s1}'$(2536)$^+ \rightarrow D^{*0} K^+$. True values of the inclusive simulation are shown. (b) is the $\bar PANDA$ full simulation of the exclusive process $\bar p p \rightarrow D_s^- D_{s0}^{*}(2317)^+$, $D_{s0}^{*}$(2317)$^+ \rightarrow D_s^+ \pi^0$. (c)  is the $\bar PANDA$ full simulation of the exclusive process $\bar p p \rightarrow D_s^-D_{s1}(2460)^+$, $D_{s1}$(2460)$^+ \rightarrow D_s^{*+} \pi^0$. All these simulations are performed with EvtGen\cite{evtgen} MC generator within the $PandaRoot$ framework. Background simulations for the process $\bar p p \rightarrow \bar q q$ with DPM\cite{dpm-panda} are ongoing and they are not reported here.}
\end{center}
\end{figure*}
\begin{figure*}[ht] 
\begin{center}
\mbox{
\subfigure[]{\scalebox{0.25}{\includegraphics{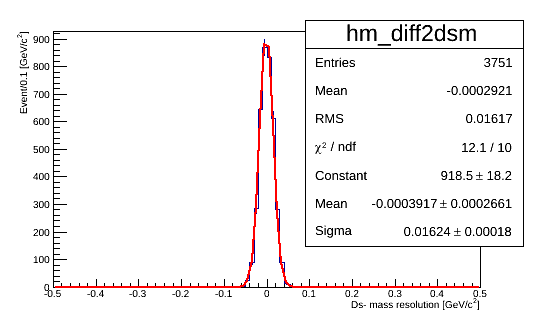}}}\quad
\subfigure[]{\scalebox{0.25}{\includegraphics{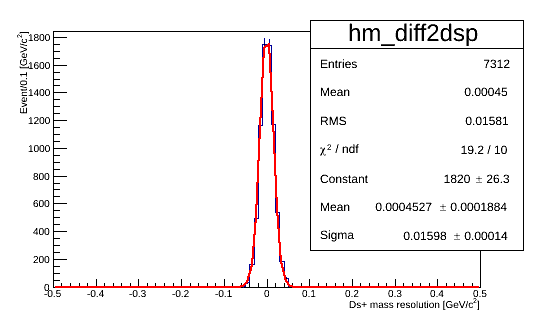}}}\quad
\subfigure[]{\scalebox{0.25}{\includegraphics{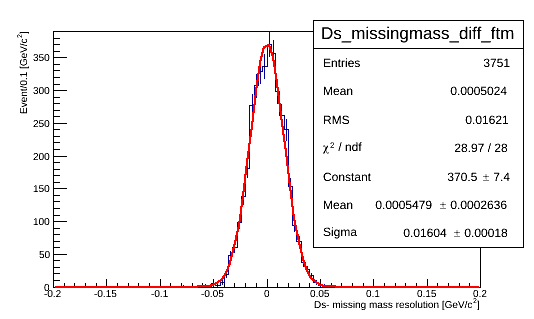}}}
}
\mbox{
\subfigure[]{\scalebox{0.25}{\includegraphics{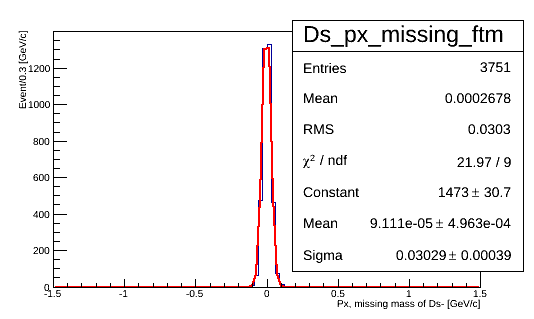}}}\quad
\subfigure[]{\scalebox{0.25}{\includegraphics{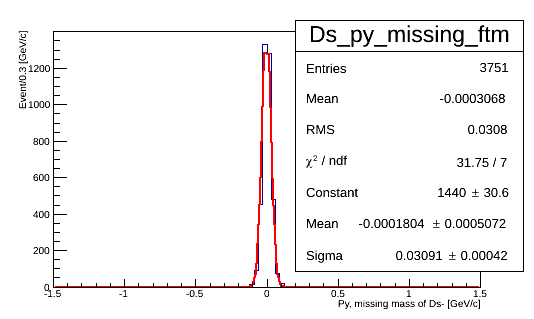}}}\quad
\subfigure[]{\scalebox{0.25}{\includegraphics{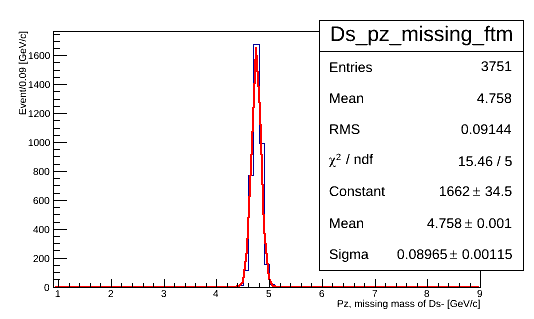}}}
}
\caption{\label{Fig40-panda} Exclusive reconstruction in $\bar PANDA$ of the physics process $\bar p p \rightarrow D_s^- D_{s0}^{*}$(2317)$^+$, with $D_{s0}^{*+}$(2317)$^+ \rightarrow D_s^+ \pi^0$ and  $D_s^\pm \rightarrow K^+ K^- \pi^\pm$. The four-momentum of $D_{s0}^{*+}$(2317)$^+$ is evaluated as missing four-momentum of the $D_s^-$ event. (a), (b) and (c) show the  mass resolution of the $D_s^-$, $D_s^+$ and the missing mass of the $D_s^-$ event (e.g., the $D_{s0}^{*}$(2317)$^+$), respectively: the mass resolution is the same in all cases, as expected. It is evaluated by subtracting the reconstructed from generated events. (d), (e) and (f) show the momentum resolution of $p_x$, $p_y$, $p_z$, respectively, e.g. the $D_{s0}^{*}$(2317)$^+$ momentum components,  reconstructed as missing momentum components of the $D_s^-$ event in this process. A vertex fit is applied in this simulations. The fit is performed by a gaussian function.}
\end{center}
\end{figure*}
The program of $\bar PANDA$ for this data analysis is:
\begin{itemize}
\item to measure the cross section of the process $\bar p p \rightarrow D_s^- D_{sJ}^{(*)+}$, where $D_{sJ}^{(*)+}$ stands for $D_{s0}^{*}$(2317)$^+$,  $D_{s1}$(2460)$^+$ and $D_{s1}'$(2536)$^+$: it is difficult to predict. What we know from experimental point of view is mostly summarized in Fig.~\ref{Fig1-panda}(a, b): these plots set up an upper and a lower limit to our expectations on the $D_s$ cross section, expected in the range of [1$-$100] nb.
\item to measure the width of the narrow states.
\item to study mixing of $D_s$ excited states with same spin (e.g. $D_{s1}^+$(2460) and $D_{s1}'$(2536)).
\item to test chiral symmetry breaking effects in the charm sector by measuring precisely mass differences, as the $D_{s0}^{*}$(2317)$^+$ and the $D_{s1}$(2460)$^+$ could be interpreted as chiral partners of the same heavy-light quark system. This involves in a precise mass measurement between the resonant state and the invariant mass system where it is observed (e.g., the mass difference between  $D_{s0}^{*}$(2317)$^+$ and $\pi^0 D_s^+$; the mass diffeence between $D_{s1}$(2460) and $\pi^0 D_s^{*+}$).
\end{itemize}
We plan to perform an inclusive (see Fig.~\ref{Fig4-panda}(a)) and an exclusive (see Fig.~\ref{Fig4-panda}(b, c)) measurement: the former for the cross section measurement, the latter for the scan of the resonance mass, indispensable to evaluate the width of the $D_{s0}^*(2317)^+$ and the $D_{s1}$(2460)$^+$, as explained later. 

\subsection{Measurement of the width of $D_{s0}^{*}$(2317)$^+$ with $\bar PANDA$}
We plan to do a mass scan in 100 keV steps for the $D_{s0}^{*}$(2317)$^+$,  $D_{s1}$(2460)$^+$ and  $D_{s1}'$(2536)$^+$. This is possible only in $\bar PANDA$, because it depends basically on the very good beam resolution and not on the detector reolution itself.

$\bar PANDA$ recent simulations, reported in Fig.~\ref{Fig4-panda} and~\ref{Fig40-panda}, are performed using the MC generator EvtGen\cite{evtgen}, within the $PandaRoot$ framework\cite{pandaroot1, pandaroot2}. We make use of same model as for the analysis described in Ref.~\cite{Dsbabar}, a Dalitz model based on real data, to simulate a realistic case and estimate the run-time needed in $\bar PANDA$ for this analysis.
The high peformance tracking detectors of $\bar PANDA$ allow excellent track reconstruction and high K/$\pi$ separation, together with the DIRC; the designed vertex detector allows high background rejection, by setting tight topological selection cuts around the fitted vertex. We reconstruct the $\rm D_s$ by using a vertex fit with three charged particles, which are identified as K or $\pi$ by means of a likelihood PID (Partile Identification)  method, that makes use of several variables like  energy loss, Cherenkov angle, Zernik momenta. Track finder and fitting procedures in the central spectrometer use the Kalman filter method; wherever the $B$ field is not homogeneous, the Runge Kutta track representation is used. In order to have better mass resolution and higher reconstruction efficiency, the  missing mass of the event is exploited: the reconstruction of the $\rm D_s^-$ is performed, then we extract information on the  $D_{s0}^*(2317)$ by evaluating its four-momentum as the difference between the reconstructed $\rm D_s^-$ four-momentum and the initial state vector. With this technique, we have obtained $\sim$30$\%$ reconstruction efficiency. A full simulation, including electronics and detector material, is being performed. In order to reject naively part of the huge background, originating from very low momentum particles, a preselection cuts on the track momentum  p$_{TRACK}>$100 MeV/c and on the photon momentum p$_{\gamma}>$50 MeV/c are applied in our simulations. A challenge of this analysis is the reconstruction of the many low momentum pions. This is the first time a full simulation including the $D_{s0}^*$(2317) is achieved with $PandaRoot$: the work in still in progress. Our simulations are based on Geant3\cite{geant3}.  

One challenge of this analysis is the background rejection, as the cross section of the process $\bar p p \rightarrow \bar q q$ is 10$^6$ times higher than the cross section expected for the signal, for $p_b>$8.8 GeV/c.  With the naive selection cuts just described, the ratio S/B went gained a factor 10$^{4}$. The optimization of the final selection criteria is now ongoing. Same approach is followed to analyze the exclusive  process $\bar p p \rightarrow D_s^- D_{s1}$(2460)$^+$: optimization of the selection to reject the huge multi-photon channel background is still ongoing (see Fig.\ref{Fig4-panda}(c)). 

\begin{figure*}[ht] 
\begin{center}
\mbox{
{\scalebox{0.4}{\includegraphics{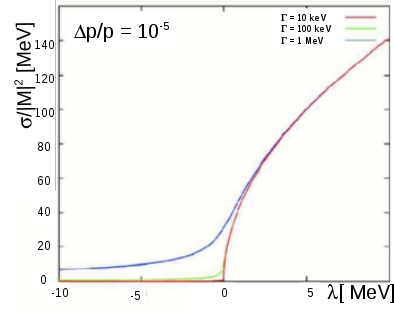}}}
}
\caption{\label{Fig5-panda} (a) Simulation in $\bar PANDA$ of the physics process $\bar p p \rightarrow D_s^- D_{s0}^{*}$(2317)$^+$, with $D_{s0}^{*}$(2317)$^+ \rightarrow D_s^+ \pi^0$ and $D_s^+ \rightarrow K^+ K^- \pi^+$. The $D_{s0}^{*}$(2317)$^+$ mass is scanned in 100 keV steps. The cross section over the matrix element $\sigma$/$\cal \lvert M \rvert ^{\rm 2}$  versus  the energy difference $\lambda$ is represented in the graph, for every point of the mass scan. The shape of the exctitation function of the cross section changes depending on the value of the width $\Gamma$ of the  $D_{s0}^{*}$(2317)$^+$, given as  input in these MC simulations.}
\end{center}
\end{figure*}

The measurement of the width  will be an original and extremely important $\bar PANDA$ contribution to solve the $cs$-spectrum puzzle. In Fig.~\ref{Fig5-panda} the plot related to the Eq.\ref{eq2-panda} is shown, scanning the $D_{s0}^{*}$(2317)$^+$ mass in 100 keV steps around its nominal value: the shape of the curve changes, depending on the input width given to the simulation.. The minimum momentum needed to produce the $D_{s0}^*$(2317)$^+$ in the process $\bar p p \rightarrow D_s^- D_{s0}^*$(2317)$^+$ is $p_b$=8.8 MeV/c; for the process $\bar p p \rightarrow D_s^- D_{s1}$(2460)$^+$ is $p_b$=9.1 MeV/c.

In high luminosity mode, e.g. with momentum resolution of the anti-proton beam $\Delta p$/$p$ = 4$\cdot$10$^{-5}$, with a cross section in the range [1$-$100] nb, and assuming a luminosity of 8.64 pb/day, with a reconstruction efficiency equal to 30$\%$, we estimate with $\bar PANDA$ a $D_{s0}^*(2317)$ production in the order of (3$-$300)$\cdot$10$^3$ events/day, to be scaled by the BF of the reconstructed $D_s$.

The $D_{s1}'(2536)$ mass is above the $D^*K$ threshold, therefore it is expected to decay dominantly to $D^{*}K$. The measurement of the width of the $D_{s1}'(2536)$ is known with high uncertainty\cite{Ds2535belle, Ds2535babar}, therefore it could be repeated in $\bar PANDA$ with higher precision and to validate our analysis thecnique. At the threshold, the Eq.~\ref{eq1-panda} becomes Eq.~\ref{eq2-panda}, so the only observables which we should measure are the mass of the resonant state, and the cross section. 

\section{Conclusion}
The Standard Model is solid, but it does not answer to all questions. Several open issues are in the sector of Charm and Charmonium physics, for example. The $\bar PANDA$ experiment at FAIR is designed to reach mass resolution 20 times better that attained at B factories, essential to perform mass scan in 100 keV steps.

In particular, in this report we put emphasis on the measurements in the sector of Charm physics with the $\bar PANDA$ experiment. These measurements represent challenges, but absolutely needed to clarify and understand the nature of these states, such as the $D_{s0}^*(2317)$ and  $D_{s1}(2460)$. $\bar PANDA$ plans to produce these narrow states at threshold; so the Coulomb term enters in the calculation: it is still not taken into account in our preliminary simulations and it represents the next step of our study.  

$\bar PANDA$ offers a unique opportunity to perform measurements with very high precision.  TDRs of several detectors are already approved, and tests with prototypes of the detectors are ongoing. The official $PandaRoot$ framework is at advanced stage to perform physics simulations. Important contributions are expected from $\bar PANDA$ when it will start to collect data.

\end{document}